\documentclass[conference]{IEEEtran}
\IEEEoverridecommandlockouts
\usepackage{cite}
\usepackage{amsmath,amssymb,amsfonts}
\usepackage{algorithm}  
\usepackage[noend]{algpseudocode}
\usepackage{graphicx}
\usepackage[caption=false,font=footnotesize]{subfig} 
\usepackage{textcomp}
\usepackage{xcolor}

\def\BibTeX{{\rm B\kern-.05em{\sc i\kern-.025em b}\kern-.08em
    T\kern-.1667em\lower.7ex\hbox{E}\kern-.125emX}}
\begin{document}

\title{Robust Beamforming for Near-Field STAR-RIS–Enabled ISCPT}

\author{
\IEEEauthorblockN{Zahra Rostamikafaki$^{1}$, Francois Chan$^{1,2}$, Claude D\textquotesingle Amours$^{1}$}
\IEEEauthorblockA{$^{1}$Department of Electrical Engineering and Computer Science, University of Ottawa, Ottawa, ON K1N 6N5, Canada}
\IEEEauthorblockA{$^{2}$Department of Electrical and Computer Engineering, Royal Military College of Canada, Kingston, ON K7K 7B4, Canada}
}

\maketitle

\begin{abstract}
A simultaneously transmitting and reflecting reconfigurable intelligent surface (STAR-RIS)–aided near-field integrated sensing, communication, and power transfer (ISCPT) framework is proposed. We formulate a robust harvested power maximization problem under imperfect cascaded  channel state information (CSI), with constraints on required user rate, eavesdropper tolerable rate, and minimum sensing beampattern gain. To address this non-convex problem, we adopt alternating optimization (AO). First, we approximate the semi-infinite inequality constraints using the S-procedure and obtain rank-one active beamforming via sequential rank-one constraint relaxation (SROCR); then we update the passive STAR-RIS coefficients with a penalty-based scheme refined by successive convex approximation (SCA). Simulations in the near field demonstrate notable gains in harvested power while meeting secrecy and beampattern targets, outperforming conventional baselines.
\end{abstract}


\section{Introduction}
Future 6G networks will natively integrate sensing, communication, and power transfer, unlocking substantial radio-resource efficiency and enabling seamless connectivity for vast fleets of low-power devices. This motivates Integrated Sensing, Communication, and Power Transfer (ISCPT), where signal design, beamforming, and resource allocation are co-optimized to sense targets and deliver information and RF energy in the same band. Compared with separate Integrated Sensing and Communication (ISAC) and Simultaneous Wireless Information and Power Transfer (SWIPT), ISCPT offers a unified framework that more efficiently balances beampattern fidelity, data-rate reliability, and harvested energy for massive IoT \cite{10663809}.

Simultaneously transmitting and reflecting RIS (STAR-RIS) is a key 6G enabler that uses low-cost reconfigurable elements to split impinging signals, enabling smart propagation control over the full 360° space \cite{10918637}. By enabling programmable transmission/reflection links, STAR-RIS can improve line-of-sight (LoS) connectivity and mitigate blockage \cite{8901159}. For large apertures and short link distances, wave propagation is spherical and must be modeled in the near field, which enables angle-and-range–dependent beamfocusing for 3D energy concentration and tighter interference control. Moreover, near-field LoS links can be rank-sufficient, supporting more spatial streams than conventional far-field LoS \cite{10716601}. 

Research on near-field ISAC or SWIPT settings remains limited \cite{10498098}, \cite{10559261}. To the best of our knowledge, no prior work has examined near-field operation in an ISCPT system particularly within a STAR-RIS framework. In near-field ISCPT, tight beamfocusing enhances not only the desired links but also the potential reception by eavesdroppers (Eves), rendering secrecy provisioning fundamentally more challenging than in the far-field regime. While robust beamforming designs for secure near-field ISAC\cite{11072251} and secure near-field SWIPT \cite{11146519} have been reported,  near-field ISCPT remains unaddressed. 

Most of the aforementioned studies \cite{10498098}, \cite{10559261} assume perfect channel state information (CSI), whereas the near-passive nature of STAR-RISs makes accurate CSI acquisition challenging \cite{9133156}.  Although robust STAR-RIS designs under CSI uncertainty have been investigated for far-field systems \cite{10304608}, the near-field case remains largely underexplored. Motivated by this gap, we develop a robust beamforming framework for STAR-RIS–aided  near-field ISCPT with imperfect CSI. The main contributions of this work are as follows:
\begin{itemize}
\item We formulate a robust near-field STAR-RIS–aided ISCPT design that maximizes the total harvested power at energy receivers (ERs), which may also act as potential eavesdroppers. The design guarantees each  information receiver (IR)’s rate requirement, limits each Eve’s worst-case rate below a secrecy threshold, and enforces a minimum sensing beampattern explicitly under bounded cascaded CSI errors.
\item We tackle the nonconvexity with an alternating-optimization (AO) framework that splits the task into two tractable subproblems: active Base Station (BS) beamforming and passive STAR-RIS phase design, solved in an iterative, decoupled manner.

\item For robustness under bounded CSI errors, we convert the semi-infinite constraints via the S-procedure, enforce rank-one active near-field beamformers using Successive Rank-One Constraint Relaxation (SROCR), and optimize the passive STAR-RIS through a penalty-based Successive Convex Approximation (SCA). Simulations validate our approach, showcasing its superiority over conventional baseline approaches.
\end{itemize}
\begin{figure}[!t]
\centering
\includegraphics[width=3.3 in]{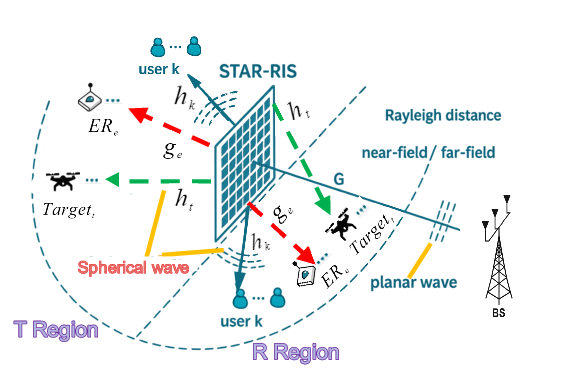}%
\caption{The Proposed ISCPT System Model.}%
\label{fig:sys}%
\end{figure}
\section{System Model and Problem Formulation}
\subsection{System Model}
We consider a STAR-RIS--assisted near-field secure ISCPT system, where an $M$-antenna BS serves multiple single-antenna IRs and ERs, while sensing several passive targets. Each ER is a potential Eve located in close proximity to its corresponding IR, attempting to intercept the legitimate information transmission. The sets of IRs, ERs, and sensing targets are denoted by $\mathcal{K}\triangleq\{1,\ldots,K\}$, $\mathcal{E}\triangleq\{1,\ldots,E\}$, and $\mathcal{T}\triangleq\{1,\ldots,T\}$, respectively.
As depicted in Fig. \ref{fig:sys}, the BS–user links are blocked by obstacles. A STAR-RIS operating in energy-splitting (ES) mode is deployed on the user side to establish LoS transmission and reflection links for the unfortunate users. The STAR-RIS employs a uniform planar array (UPA) of $N = N_y \times N_z$ elements, while the BS is equipped with uniform linear arrays (ULAs). Without loss of generality, users are randomly located in both the reflection (R) and transmission (T) regions of the STAR-RIS, corresponding to $l=r$ and $l=t$, respectively. Each STAR-RIS element $n\in\{1,\ldots,N\}$ simultaneously
transmits and reflects the incident signal $x_n$ from the BS, modeled as $z_n^{l}=\sqrt{\beta_n^{\,l}}\,\mathrm{e}^{\jmath \theta_{n}^l}\,x_n$, with $ l\in\{t,r\}$,  $\beta_{n}^l\in[0,1]$, and $\theta_{n}^l\in[0,2\pi)$. The transmission–reflection coefficients (TRCs) of the STAR-RIS are collected in $\boldsymbol{\Theta}_l=\operatorname{diag}\!\big(\sqrt{\beta_{1}^{\,l}}\,e^{\mathrm{j}\theta_{1}^{\,l}},\ldots,\sqrt{\beta_{N}^{\,l}}\,e^{\mathrm{j}\theta_{N}^{\,l}}\big)
$, and element-wise energy conservation requires $\beta_{n}^t+\beta_{n}^r=1$.
\subsection{Channel Model}
Given the large separation between the BS and the STAR-RIS, a far-field model is adopted for the BS--STAR-RIS link, whereas the proximity of IRs, ERs, and dedicated targets to the STAR-RIS motivates a near-field model for the STAR-RIS-receiver links. The far-field channel from the BS to the STAR-RIS, and the near-field channels from the STAR-RIS to the $k$-th IR, $e$-th ER, and $t$-th target are denoted by $\mathbf{G}\in\mathbb{C}^{N\times M}$, $\mathbf{h}_k\in\mathbb{C}^{N\times 1}$,  $\mathbf{g}_e\in\mathbb{C}^{N\times 1}$, and $\mathbf{h}_{t}\in\mathbb{C}^{N\times 1}$ respectively. For the far-field BS--STAR-RIS link, we adopt the widely used $L$-scatterer geometric channel model, given by \cite{10437418}
\begin{equation}
\mathbf{G}
= \sqrt{\frac{\nu\,M N}{L}}
\sum_{\ell_s=1}^{L}
\mathbf{a}_{\mathrm{STAR}}(\vartheta_{\ell_s},\psi_{\ell_s})\,
\mathbf{a}_{\mathrm{BS}}^{H}(\varphi_{\ell_s}),
\label{eq:1}
\end{equation}
where $\varphi_{\ell_s}$ is the BS angle of departure (AoD), $(\vartheta_{\ell_s},\psi_{\ell_s})$ are the azimuth and elevation angles of arrivals (AoAs) at the STAR-RIS, $L$ is the number of dominant paths, and $\nu$ is the large-scale path-loss factor. The steering vectors $\mathbf{a}_{\mathrm{STAR}}(\cdot)$ and $\mathbf{a}_{\mathrm{BS}}(\cdot)$ correspond to the UPA at the STAR-RIS and the ULA at the BS, respectively.

For the near-field geometry, the STAR-RIS is located on the $YZ$-plane and receiving nodes and sensing targets are located on the $XY$-plane. Let the reference STAR-RIS element be at $(0,y_{ref},z_{ref})$. If the aperture is indexed row by row from the bottom to the top, the Cartesian coordinate of element $n\in\{1,\ldots,N\}$ is $
\mathbf{s}_n=\big[\,0,\; y_{ref} + i_y(n)\,d_S,\; z_{ref} + i_z(n)\,d_S\,\big]^{T}$ where the inter-element spacing is $d_S=\lambda_c$ \cite{10437418}, the column index is $i_y(n)=\operatorname{mod}(n-1,N_y)$ and row index is $i_z(n)=\left\lfloor\frac{n-1}{N_y}\right\rfloor$. IRs, ERs, and sensing targets are located at
$\mathbf{p}_k^{\mathrm{I}}=[x_k,y_k,0]^{T}$, $\mathbf{p}_e^{\mathrm{E}}=[x_{e},y_{e},0]^{T}$, and $\mathbf{p}_t^{\mathrm{T}}=[x_{t},y_{t},0]^{T}$, respectively. 

Adopting a unified spherical-wave propagation model, the LoS channel between the $n$-th STAR element and a point located at $\mathbf{p} \in \left\{ \mathbf{p}_k^{\mathrm{I}},\, \mathbf{p}_e^{\mathrm{E}},\, \mathbf{p}_t^{\mathrm{T}} \right\}$ is given element-wise in \cite{10437418} as
\begin{equation}
[\mathbf{b}(\mathbf{p})]_n
= \frac{\lambda_c}{4\pi \,\lVert \mathbf{p}-\mathbf{s}_n\rVert_{2}}
\exp\!\left(-\mathrm{j}\,\frac{2\pi}{\lambda_c}\,\lVert \mathbf{p}-\mathbf{s}_n\rVert_2\right).
\label{eq:2}
\end{equation}
Accordingly, $\mathbf{h}_k=\mathbf{b}(\mathbf{p}_k^{\mathrm{I}})$, $\mathbf{g}_e=\mathbf{b}(\mathbf{p}_e^{\mathrm{E}})$, and $\mathbf{h}_{t}=\mathbf{b}(\mathbf{p}_t^{\mathrm{T}})$ denote the STAR-RIS channels associated with an IR, an ER, and a target, respectively.

\subsection{Signal Transmission Model}
The signal sent by the BS can be expressed as \cite{11072251}
\begin{equation}\label{eq:3}
\mathbf{x}=\sum_{k \in \mathcal{K}} \mathbf{w}_k s_k + \mathbf{v},
\end{equation}
where the information symbols $s_k$, assumed independent across users with unit power $\mathbb{E}\{|s_k|^2\}=1$, $\forall\,k \in \mathcal{K}$, $\mathbf{w}_k\in\mathbb{C}^{M\times 1}$ beamforming vector for user $k$, and $\mathbf{v}\in\mathbb{C}^{M\times 1}$ is the dedicated sensing waveform independent of $\{s_k\}$ with covariance $\mathbf{V}=\mathbb{E}\{\mathbf{v}\mathbf{v}^{H}\}\succeq \mathbf{0}$. 

Then, the received signal at IR$_k$ and ER$_e$, $\forall\,k \in \mathcal{K}$, $\forall\,e \in \mathcal{E}$ are modeled as
\begin{equation}\label{eq:4}
y_{k}=\mathbf{h}_k^{H}\boldsymbol{\Theta}_l\mathbf{G}\mathbf{x}+n_{k},
\end{equation}
\begin{equation}\label{eq:5}
y_{e,k}=\mathbf{g}_e^{H}\boldsymbol{\Theta}_l\mathbf{G}\mathbf{x}+n_{e},
\end{equation}
with $n_{k}\sim \mathcal{CN}(0,\sigma_{I}^{2})$ and $n_{e}\sim \mathcal{CN}(0,\sigma_{e}^{2})$ denote the additive white Gaussian noise at IR$_k$ and ER$_e$ respectively. 

Accordingly, the achievable communication rate received at user $k$ on side $\l\in\{t,r\}$ is given by
\begin{equation}\label{eq:6}
R_{k}^{IR} = \log_2\!\left( 1 +
\frac{ \left|\boldsymbol{\Phi}_l^{H}\mathbf{H}_k\mathbf{w}_k\right|^{2} }
{ \sum_{\substack{k'=1\\k'\neq k}}^{K} \left|\boldsymbol{\Phi}_l^{H}\mathbf{H}_k\mathbf{w}_{k'}\right|^{2}
  + \left|\boldsymbol{\Phi}_l^{H}\mathbf{H}_k\mathbf{v}\right|^{2}
  + \sigma_{I}^{2} }
\right),
\end{equation}
where $\boldsymbol{\Phi}_l = \operatorname{diag}(\mathbf{\Theta_{l}})$, and $\mathbf{H}_k=\operatorname{diag}(\mathbf{h}_k)\mathbf{G}$ denotes the cascaded channel from the BS to user $k$.

Assuming an ER attempts to intercept the confidential message intended for an IR, the achievable eavesdropping rate can be expressed as
\begin{equation}\label{eq:7}
R_{e,k}^{Eve}=\log_2\!\left( 1 +
\frac{\left|\boldsymbol{\Phi}_l^{H}\mathbf{F}_e\mathbf{w}_k\right|^{2}}
{ \sum_{\substack{k'=1\\k'\neq k}}^{K}\left|\boldsymbol{\Phi}_l^{H}\mathbf{F}_e\mathbf{w}_{k'}\right|^{2}
  +\left|\boldsymbol{\Phi}_l^{H}\mathbf{F}_e\mathbf{v}\right|^{2}
  + \sigma_{e}^{2} }\right),
\end{equation}
where  $ l\in\{t,r\}$, and $\mathbf{F}_{e}=\operatorname{diag}(\mathbf{g}_{e})\mathbf{G}$ represents the cascade channel from BS to ER$_e$.

At the energy receiver, interference and noise are typically negligible \cite{10304608}. Thus, the received RF power is given by 
\begin{equation}\label{eq:8}
E_{e}=\eta_e\,\mathbb{E}\!\left\{\left|\boldsymbol{\Phi}_l^{H}\mathbf{F}_e\mathbf{x}\right|^{2}\right\}
,  \quad l\in\{t,r\}
\end{equation}
where $\eta_e\in(0,1]$ is the energy-harvesting efficiency factor.

Sensing performance is quantified by the beampattern gain, as greater illumination power at each target improves detection reliability. For target $t$, the instantaneous sensing power is
\begin{equation}\label{eq:9}
\mathcal{P}_t^{\mathrm{tar}}=\mathbb{E}\!\left\{\left|\mathbf{\Phi}_l^{H}\mathbf{H}_t\mathbf{x}\right|^{2}\right\}
, \quad l\in\{t,r\}
\end{equation}
where $\mathbf{H}_{t}=\operatorname{diag}(\mathbf{h}_{t})\mathbf{G}$ is the effective channel from the BS via the STAR-RIS to the target on side $ l\in\{t,r\}$. 

Owing to the passive aperture and channel complexity, perfect CSI for the links between the BS and the STAR-RIS and for the links from the STAR-RIS to the users and targets is impractical. We therefore adopt a bounded-error model on the cascaded channels. For IR$_k$, let $\mathbf{H}_k=\hat{\mathbf{H}}_k+\Delta\mathbf{H}_k$ with uncertainty set $\mathcal{H}_k=\{\Delta\mathbf{H}_k\in\mathbb{C}^{N\times M}:\,\|\Delta\mathbf{H}_k\|_F\le\delta_{k}\}$; for ER$_e$, let $\mathbf{F}_e=\hat{\mathbf{F}}_e+\Delta\mathbf{F}_e$ with $\mathcal{F}_e=\{\Delta\mathbf{F}_e\in\mathbb{C}^{N\times M}:\,\|\Delta\mathbf{F}_e\|_F\le\delta_{e}\}$. For $t$-th target,  $\mathbf{H}_{t}=\hat{\mathbf{H}}_{t}+\Delta\mathbf{H}_{t}$ with $\mathcal{H}_{t}=\{\Delta\mathbf{H}_{t}\in\mathbb{C}^{N\times M}:\,\|\Delta\mathbf{H}_{t}\|_F\le\delta_{t}\}$. Here $\hat{\mathbf{H}}_k$,  $\hat{\mathbf{F}}_e$, and $\hat{\mathbf{H}}_{t}$ are the estimated cascaded channels, while $\Delta\mathbf{H}_k$, $\Delta\mathbf{F}_e$, and $\Delta\mathbf{H}_{t}$ denote the estimation errors bounded in Frobenius norm by $\delta_{k}$, $\delta_{e}$, and $\delta_{t}$  respectively.
\subsection{Problem Formulation}
In this paper, we aim to maximize the total harvested power across all ERs while guaranteeing the communication rate for each $\mathrm{IR}_k$ and satisfying a sensing beampattern requirement, by jointly optimizing the active beamformers $\{\mathbf{w}_k\}$, the sensing covariance matrix $\mathbf{V}$, and the passive STAR-RIS coefficients $\{\boldsymbol{\Phi}_l\}$. Under imperfect CSI, the resulting robust design is formulated as:
\begin{subequations}
\begin{align}
\max_{\{\mathbf{w}_k\},\,\{\mathbf{V}\ge 0\},\,\{\mathbf{\Phi}_l\}} \quad
& \sum_{e \in \mathcal{E}} E_{e} \tag{\theparentequation}\label{eq:10}\\
\text{s.t.}\quad 
& R_{k}^{IR} \ge R_{\mathrm{th}}, \quad \forall\, k  \label{eq:10a}\\
& R_{e,k}^{Eve} \le R_{e,\mathrm{th}}, \quad \forall\, k , \forall\, e \label{eq:10b}\\
& \mathcal{P}_t^{\mathrm{tar}} \ge \Lambda, \quad \forall\, t\label{eq:10c}\\
& \sum_{k=1}^{K}\lVert \mathbf{w}_k \rVert_2^2 + \operatorname{tr}(\mathbf{V}) \le P_{\max} \label{eq:10d}\\
& \theta_{n}^l \in [0,2\pi),  \forall n\in {\mathcal{N}},\ l\in\{t,r\};\label{eq:10e}\\ 
& \beta_n^{t},\beta_n^{r}\in[0,1];
\beta_n^{t}+\beta_n^{r}=1,\ \forall n\label{eq:10f}
\end{align}
\end{subequations}
where $R_{\mathrm{th}}$ denotes the minimum rate required at the information receivers, while $R_{\mathrm{e,th}}$ bounds the eavesdroppers’ worst-case rate to guarantee secrecy for IR$_k$. Constraint \eqref{eq:10c} enforces a minimum sensing beampattern gain for target detection. Constraint \eqref{eq:10d} represents the total power budget at BS. \eqref{eq:10e} and \eqref{eq:10f} are the constraints on phase and amplitude of the STAR-RIS elements respectively.

Note that problem \eqref{eq:10} is a nonconvex program and cannot be solved directly. 
The difficulty arises from (i) the strong coupling among the decision variables 
$\{\mathbf{w}_k\}$, $\mathbf{V}$, and $\{\boldsymbol{\Phi}_l\}$, and 
(ii) CSI uncertainty, which renders the objective and constraints \eqref{eq:10a}, \eqref{eq:10b}, and \eqref{eq:10c} semi-infinite. Consequently, we develop an efficient alternating-optimization (AO) 
procedure to obtain a high-quality suboptimal solution.

\section{Proposed Solution}
In this section, problem \eqref{eq:10} is first recast as a rank-constrained semidefinite program (SDP). Leveraging the S-procedure, the resulting semi-infinite constraints are converted into tractable linear matrix inequalities (LMIs). The reformulated model is then split into two subproblems, in which the active and passive beamformers are optimized in an alternating manner.
\subsection{Problem Reformulation}
To proceed, we introduce the lifted variables $\mathbf{W}_k=\mathbf{w}_k\mathbf{w}_k^{H}$, so that $\mathbf{W}_k\succeq\mathbf{0}$ and $\mathrm{rank}(\mathbf{W}_k)=1$. 
Moreover, for the STAR-RIS we define $\mathbf{Q}_{l}=\boldsymbol{\Phi}_{l}\boldsymbol{\Phi}_{l}^{H}$ with $\l\in\{t,r\}$, thereby enforcing $\mathbf{Q}_{l}\succeq\mathbf{0}$, $\mathrm{rank}(\mathbf{Q}_{l})=1$, and $\mathrm{diag}(\mathbf{Q}_{l})=\boldsymbol{\beta}_{l}$ where $\boldsymbol{\beta}_{l}=[\beta_{1}^{l},\ldots,\beta_{N}^{l}]^{\top}$. In addition, we introduce an auxiliary scalar $\xi $ so that $\sum_e E_{e}\ge \xi$ holds for all $e\in\mathcal{E}$. Consequently, with these constructions, problem \eqref{eq:10} is recast as 
\begin{subequations}
\begin{align}
& \max_{\{\mathbf{W}_k\}, \{\mathbf{V} \succeq \mathbf{0}\}, \{\mathbf{Q}_l\}, \xi} \quad \xi \tag{\theparentequation}\label{eq:11} \\
& \text{s.t.} \quad \text{Tr}\left( \left(\sum_{k=1}^K \mathbf{W}_k + \mathbf{V}\right) \mathbf{F}_e^H\mathbf{Q}_l\mathbf{F}_e \right) \geq \xi, \notag \\ 
& \qquad \quad  l \in \{t, r\}, \quad \forall\, e\label{eq:11a} \\
& \qquad \quad \text{Tr}\left( \left(\frac{\mathbf{W}_k}{\Gamma} - \sum_{\substack{k'=1\\ k'\neq k}}^{K} \mathbf{W}_{k'} - \mathbf{V}\right) \mathbf{H}_k^H \mathbf{Q}_l\mathbf{H}_k\right) \geq \sigma_{I}^2,  \notag \\
& \qquad \quad  l \in \{t, r\}, \quad \forall\, k, \label{eq:11b} \\
& \qquad \quad \text{Tr}\left( \left(\sum_{\substack{k'=1\\ k'\neq k}}^{K} \mathbf{W}_{k'} + \mathbf{V} - \frac{\mathbf{W}_k}{\eta}\right) \mathbf{F}_e^H\mathbf{Q}_l\mathbf{F}_e \right) + \sigma_{e}^2 \geq 0,  \notag \\
& \qquad \quad  l \in \{t, r\}, \quad \forall\, k,  \forall\, e,\label{eq:11c} \\
& \qquad \quad \text{Tr}\left( \left(\sum_{k=1}^K \mathbf{W}_k + \mathbf{V}\right) \mathbf{H}_{t}^H\mathbf{Q}_l\mathbf{H}_{t} \right) \geq \Lambda, \notag \\
& \qquad \quad  l \in \{t, r\}, \quad \forall\, t,\label{eq:11d} \\
& \qquad \quad \sum_{k=1}^K \text{Tr}(\mathbf{W}_k) + \text{Tr}(\mathbf{V}) \leq P_{\text{max}}, \label{eq:11e} \\
& \qquad \quad  \mathbf{W}_k \succeq 0, \text{Rank}(\mathbf{W}_k) = 1,\label{eq:11f}\\
& \qquad \quad \mathbf{Q}_l \succeq 0, \text{Rank}(\mathbf{Q}_l) = 1.  \quad l \in \{t, r\} \label{eq:11g}\\
& \qquad \quad \mathrm{diag}(\mathbf{Q}_l) = \boldsymbol{\beta}_l, \quad \forall l \label{eq:11h}\\
& \qquad \quad \beta^t_n + \beta^r_n = 1, 0 \leq \beta^{t/r}_n \leq 1
\quad \forall n \in \mathcal{N}. \label{eq:11i}
\end{align}
\end{subequations}
where $\Gamma = 2^{R_{\mathrm{th}}}-1$ and $\eta = 2^{R_{\mathrm{e,th}}}-1$ denote the minimum  signal-to-interference-plus-noise ratio (SINR) required at the IRs and the maximum SINR allowed at the eavesdroppers, respectively.

Exploiting the cyclic property of the trace, $\mathrm{Tr}(\mathbf{ABCD})=\mathrm{Tr}(\mathbf{DABC})$, together with the vectorization identity $\mathrm{Tr}(\mathbf{A}^{H}\mathbf{B}\mathbf{C}\mathbf{D})=\mathrm{vec}(\mathbf{A})^{H}\!\left(\mathbf{D}^{T}\!\otimes\!\mathbf{B}\right)\mathrm{vec}(\mathbf{C})$ \cite{10304608}, the constraints in \eqref{eq:11a}–\eqref{eq:11d} can be formulated as:
\begin{equation}
\label{eq:12}
\mathrm{vec}(\mathbf{F}_e)^{H}
\Big( \big( \textstyle\sum_{k=1}^{K}\mathbf{W}_k+\mathbf{V} \big)^{T} \otimes \mathbf{Q}_{l} \Big)
\mathrm{vec}(\mathbf{F}_e) \;\ge\; \xi
\end{equation}
\begin{equation}
\label{eq:13}
\mathrm{vec}(\mathbf{H}_k)^{H}
\Big( \big( \tfrac{\mathbf{W}_k}{\Gamma}-\!\!\sum_{\substack{k'=1\\ k'\neq k}}^{K}\mathbf{W}_{k'}-\mathbf{V} \big)^{T} \otimes \mathbf{Q}_{l} \Big)
\mathrm{vec}(\mathbf{H}_k) \;\ge\; \sigma_{I}^{2}
\end{equation}
\begin{equation}
\mathrm{vec}(\mathbf{F}_{e})^{H}
\Big( \big( \sum_{\substack{k'=1\\ k'\neq k}}^{K}\mathbf{W}_{k'} + \mathbf{V} - \tfrac{\mathbf{W}_{k}}{\eta} \big)^{\!T} \otimes \mathbf{Q}_{l} \Big)
\mathrm{vec}(\mathbf{F}_{e}) \;+\; \sigma_{e}^{2} \;\ge\; 0,
\label{eq:14}
\end{equation}
\begin{equation}
\mathrm{vec}(\mathbf{H}_{t})^{H}
\Big( \big( \sum_{k=1}^{K}\mathbf{W}_{k} + \mathbf{V} \big)^{\!T} \otimes \mathbf{Q}_{l} \Big)
\mathrm{vec}(\mathbf{H}_{t}) \;\ge\; \Lambda .
\label{eq:15}
\end{equation}

To account for the channel uncertainties in \eqref{eq:12}-\eqref{eq:15}, we invoke the general S-procedure \cite{boyd2004convex} directly on the quadratic forms obtained after vectorization. 
Let the channel in \eqref{eq:13} be expressed as $\mathbf{H}_k = \hat{\mathbf{H}}_k + \Delta \mathbf{H}_k$ and $\mathbf{S}_1 = (\tfrac{\mathbf{W}_k}{\Gamma}-\!\!\sum_{\substack{k'=1\\ k'\neq k}}^{K}\mathbf{W}_{k'}-\mathbf{V} \big)^{T}$. 
By substituting into the constraint, we have
\begin{equation}
\operatorname{vec}(\hat{\mathbf{H}}_k + \Delta \mathbf{H}_k)^H 
\left( \mathbf{S}_1\otimes \mathbf{Q}_{l} \right)
\operatorname{vec}(\hat{\mathbf{H}}_k + \Delta \mathbf{H}_k) - \sigma_{I}^2 \geq 0,
\label{eq:16}
\end{equation}
which is equivalent to
\begin{equation}
\operatorname{vec}(\Delta \mathbf{H}_k)^H \mathbf{A}_1 \operatorname{vec}(\Delta \mathbf{H}_k) 
+ 2 \Re\left\{ \operatorname{vec}(\Delta \mathbf{H}_k)^H \mathbf{b} \right\} + c \geq 0,
\label{eq:17}
\end{equation}
where, we define $\mathbf{A}_1 = \mathbf{S}_1 \otimes \mathbf{Q}_{l}$, 
$\mathbf{b} = \mathbf{A}_1 \operatorname{vec}(\hat{\mathbf{H}}_k)$, 
and $c = \operatorname{vec}(\hat{\mathbf{H}}_k)^H \mathbf{b} - \sigma_{I}^2$.

Since $\|\Delta \mathbf{H}_k\|_{F}\le \delta_{k}$ implies $\|\mathrm{vec}(\Delta \mathbf{H}_k)\|_{2}\le \delta_{k}$, the channel uncertainty set can be equivalently written as the quadratic constraint
\begin{equation} \label{eq:18}
-\mathrm{vec}(\Delta \mathbf{H}_k)^{H}\mathbf{I}\,\mathrm{vec}(\Delta \mathbf{H}_k)+\delta_{k}^{2}\ \ge\ 0,
\end{equation}

By applying the S-procedure, \eqref{eq:17} together with \eqref{eq:18} yields the following linear matrix inequality.

\begin{equation} \label{eq:19}
\begin{pmatrix}
\mathbf{S}_1 \otimes \mathbf{Q}_l + \tau_{k} \mathbf{I} & (\mathbf{S}_1 \otimes \mathbf{Q}_l)\, \operatorname{vec}(\hat{\mathbf{H}}_k) \\
\operatorname{vec}(\hat{\mathbf{H}}_k)^H\, (\mathbf{S}_1 \otimes \mathbf{Q}_l) & \zeta_k
\end{pmatrix}
\succeq 0
\end{equation}
where \( \zeta_k = \operatorname{vec}(\hat{\mathbf{H}}_k)^H (\mathbf{S}_1 \otimes \mathbf{Q}_l) \operatorname{vec}(\hat{\mathbf{H}}_k) - \sigma_{I}^2 - \tau_{k}  \delta_{k}^2 \), \( \tau_{k} \geq 0 \) is an auxiliary variable introduced by the S-procedure, and \( \delta_{k} \) denotes the uncertainty bound.

Likewise, constraints \eqref{eq:12}, \eqref{eq:14}, and \eqref{eq:15} are cast as LMIs: 
 
\begin{equation} \label{eq:20}
\begin{pmatrix}
\mathbf{S}_{2}\otimes \mathbf{Q}_l + \tau_{e} \mathbf{I} & (\mathbf{S}_{2}\otimes \mathbf{Q}_l) \operatorname{vec}(\hat{\mathbf{F}}_e) \\
\operatorname{vec}(\hat{\mathbf{F}}_e)^H (\mathbf{S}_{2}\otimes \mathbf{Q}_l) & \mu_e
\end{pmatrix}
\succeq 0,
\end{equation}
where  $\mathbf{S}_{2} = \big( \textstyle\sum_{k=1}^{K}\mathbf{W}_k+\mathbf{V} \big)^{T}$, $\mu_e = \operatorname{vec}(\hat{\mathbf{F}}_e)^H (\mathbf{S}_{2}\otimes \mathbf{Q}_l) \operatorname{vec}(\hat{\mathbf{F}}_e) - \xi - \tau_{e} \delta_{e}^2$, and $\tau_{e} \geq 0$.

\begin{equation} \label{eq:21}
\begin{pmatrix}
\mathbf{S}_{3}\otimes \mathbf{Q}_l + \tau_{e,k} \mathbf{I} & (\mathbf{S}_{3}\otimes \mathbf{Q}_l) \operatorname{vec}(\hat{\mathbf{F}}_e) \\
\operatorname{vec}(\hat{\mathbf{F}}_e)^H (\mathbf{S}_{3}\otimes \mathbf{Q}_l) & \varrho_{e,k}
\end{pmatrix}
\succeq 0,
\end{equation}
where  $\mathbf{S}_{3} = \big( \sum_{\substack{k'=1\\ k'\neq k}}^{K}\mathbf{W}_{k'} + \mathbf{V} - \tfrac{\mathbf{W}_{k}}{\eta} \big)^{\!T}$, $\varrho_{e,k} = \operatorname{vec}(\hat{\mathbf{F}}_e)^H (\mathbf{S}_{3}\otimes \mathbf{Q}_l) \operatorname{vec}(\hat{\mathbf{F}}_e) + \sigma_{e}^{2} - \tau_{e,k} \delta_{e}^2$, and $\tau_{e,k} \geq 0$.

\begin{equation} \label{eq:22}
\begin{pmatrix}
\mathbf{S}_{4}\otimes \mathbf{Q}_l + \tau_{t} \mathbf{I} & (\mathbf{S}_{4}\otimes \mathbf{Q}_l) \operatorname{vec}(\hat{\mathbf{H}}_{t}) \\
\operatorname{vec}(\hat{\mathbf{H}}_{t})^H (\mathbf{S}_{4}\otimes \mathbf{Q}_l) & \varsigma_t
\end{pmatrix}
\succeq 0,
\end{equation}
where  $\mathbf{S}_{4} = \big( \textstyle\sum_{k=1}^{K}\mathbf{W}_k+\mathbf{V} \big)^{T}$, $\varsigma_t = \operatorname{vec}(\hat{\mathbf{H}}_{t})^H (\mathbf{S}_{4}\otimes \mathbf{Q}_l) \operatorname{vec}(\hat{\mathbf{H}}_{t}) - \Lambda - \tau_{t} \delta_{t}^2$, and $\tau_{t} \geq 0$.

After these transformations, problem \eqref{eq:11} is recast as:
\begin{subequations} 
\begin{align}
\max_{\{\tau\}, \{\xi\}, \{\mathbf{Q}_l\}, \{\mathbf{V}\succeq \mathbf{0}\}, \{\mathbf{W}_k\}} \quad
\xi \tag{\theparentequation}\label{eq:23}\\
\text{s.t.}\quad
\eqref{eq:11e}- \eqref{eq:11i}, \eqref{eq:19}-\eqref{eq:22}\label{eq:23a}.
\end{align}
\end{subequations}
where $\{\tau\}$ denotes the set of S-procedure multipliers $\{\tau_{k},\tau_{e},\tau_{e,k},\tau_{t}\}$. Since the variables in \eqref{eq:19}-\eqref{eq:22} remain coupled, we decompose problem \eqref{eq:23} into active and passive beamforming subproblems, which are solved alternately within our AO framework.
\subsection{Active Beamforming Design}
We now focus on the active beamforming subproblem. For a given ${\mathbf{Q_l}}$, problem \eqref{eq:23} is rewritten as:
\begin{subequations} 
\begin{align}
\max_{\{\tau\}, \{\xi\}, \{\mathbf{V}\succeq \mathbf{0}\}, \{\mathbf{W}_k\}} \quad & \xi \tag{\theparentequation}\label{eq:24} \\
\text{s.t.} \quad
\eqref{eq:11e}, \eqref{eq:11f}, \eqref{eq:19}-\eqref{eq:22}\label{eq:24a}.
\end{align}
\end{subequations}

The non-convexity of problem \eqref{eq:24} arises from the rank-one constraints \eqref{eq:11f}. We address this using the SROCR method. First, the non-convex rank-one constraint in \eqref{eq:11f}  can be equivalently expressed as
\begin{equation}\label{eq:25}
\kappa_{\max}(\mathbf{W}_k) = \mathrm{tr}(\mathbf{W}_k),\quad \forall k \in \mathcal{K},
\end{equation}
where $\kappa_{\max}(\cdot)$ denotes the largest eigenvalue. Following SROCR algorithm \cite{CaoEUSIPCO2017}, during $i$-th iteration we introduce a relaxation parameter $u_k^{(i-1)} \in [0,1]$ for each rank-one constraint $\mathrm{rank}(\mathbf{W}_k)=1$ and enforce
\begin{equation} \label{eq:26}
\big(\mathbf{q}_{k,\max}^{(i-1)}\big)^{\!H}\,\mathbf{W}_k\,\mathbf{q}_{k,\max}^{(i-1)}
\;\ge\;
u_k^{(i-1)}\,\mathrm{tr}(\mathbf{W}_k),
\quad \forall k \in \mathcal{K},
\end{equation}
where $\mathbf{q}_{k,\max}^{(i-1)}$ is the dominant eigenvector of $\mathbf{W}_k^{(i-1)}$ obtained in the previous iteration. With these relaxations, each SROCR step reduces to solving a convex SDP. To implement SROCR for \eqref{eq:24}, at iteration $i$ we solve the following SDP:
\begin{subequations}
\begin{align}
\max_{\{\tau\}, \{\xi\}, \{\mathbf{V}\succeq \mathbf{0}\}, \{\mathbf{W}_k\}} \quad & \xi \tag{\theparentequation}\label{eq:27} \\
\text{s.t.}\quad
\eqref{eq:11e}, \eqref{eq:19}-\eqref{eq:22} \label{eq:27a} \\
\big(\mathbf{q}_{k,\max}^{(i-1)}\big)^{\!H}\,\mathbf{W}_k\,\mathbf{q}_{k,\max}^{(i-1)}
\;\ge\;
u_k^{(i-1)}\,\mathrm{tr}\!\big(\mathbf{W}_k\big), \notag \\
\forall k \in \mathcal{K}. \label{eq:27b}
\end{align}
\end{subequations}
which is convex and amenable to efficient solution using standard convex-optimization solvers.

\subsection{Passive Beamforming Design}
With $\{\mathbf{W}_k\}$ and $\{\mathbf{V}\}$ fixed, we optimize the passive beamforming. The subproblem reduces to
\begin{subequations} 
\begin{align}
\max_{\{\tau\}, \{\xi\}, \{\mathbf{Q}_l\}, } \quad & \xi \tag{\theparentequation}\label{eq:28} \\
\text{s.t.} \quad
\eqref{eq:11g}-\eqref{eq:11i}, \eqref{eq:19}-\eqref{eq:22}\label{eq:28a}.
\end{align}
\end{subequations}

As with \eqref{eq:24} , the rank-one constraint in \eqref{eq:11g} precludes a direct optimal solution due to its non-convex nature. Therefore, we reformulate this constraint into an equivalent form, as established in \cite{10304608}:
\begin{equation}
\label{eq:29}
\|\mathbf{Q}_{l}\|_{*}-\|\mathbf{Q}_{l}\|_{2}=0,\quad  \l\in\{t,r\},
\end{equation}
where $\|\mathbf{Q}_{l}\|_{*}$ and $\|\mathbf{Q}_{l}\|_{2}$ denote the nuclear and spectral norms of $\mathbf{Q}_{l}$, respectively. Because $\mathbf{Q}_{l}\succeq \mathbf{0}$, one always has $\|\mathbf{Q}_{l}\|_{*}-\|\mathbf{Q}_{l}\|_{2}\ge 0$, with equality if and only if $\operatorname{rank}(\mathbf{Q}_{l})=1$. By adopting the penalty method, we incorporate the equivalent rank equation as a penalty term in the objective function. Consequently, problem \eqref{eq:28} can be reformulated as: 
\begin{subequations} 
\begin{align}
\max_{\{\tau\},\{\mathbf{Q}_{l}\},\{\xi\}}\;
\xi-\gamma\sum_{l\in\{t,r\}}\Big(\|\mathbf{Q}_{l}\|_{*}-\|\mathbf{Q}_{l}\|_{2}\Big) \tag{\theparentequation}\label{eq:30}\\
\text{s.t. }\quad
\eqref{eq:11h}, \eqref{eq:11i}, \eqref{eq:19}-\eqref{eq:22}  \label{eq:30a}.
\end{align}
\end{subequations}
where $\gamma>0$ is a penalty factor that penalizes the objective function when the rank of $\mathbf{Q_l}$ deviates from one. The resulting problem \eqref{eq:30} is still non-convex on account of the penalty term. Therefore, we leverage SCA to obtain a convex surrogate function by applying the first-order Taylor approximation, which yields the following upper bound:
\begin{equation}
\label{eq:31}
\|\mathbf{Q}_{l}\|_{*}-\|\mathbf{Q}_{l}\|_{2}\;\le\;\|\mathbf{Q}_{l}\|_{*}-\widehat{\mathbf{Q}}_{l}^{(j)},
\end{equation}
with $\widehat{\mathbf{Q}}_{l}^{(j)}=\|\mathbf{Q}_{l}^{(j)}\|_{2}+\operatorname{Tr}\!\Big(\mathbf{\varkappa}(\mathbf{Q}_{l}^{(j)})\,\mathbf{\varkappa}(\mathbf{Q}_{l}^{(j)})^{H}\big(\mathbf{Q}_{l}-\mathbf{Q}_{l}^{(j)}\big)\Big)$,
where $\mathbf{\varkappa}(\mathbf{Q}_{l}^{(j)})$ is the dominant eigenvector of $\mathbf{Q}_{l}^{(j)}$. Using this surrogate, the problem at iteration $j$ becomes
\begin{subequations} 
\begin{align}
\max_{\{\tau\},\{\mathbf{Q}_{l}\},\{\xi\}}\;
\xi-\gamma\sum_{l\in\{t,r\}}\Big(\|\mathbf{Q}_{l}\|_{*}-\widehat{\mathbf{Q}}_{l}^{(j)}\Big) \tag{\theparentequation}\label{eq:32}\\
\text{s.t. }\quad
\eqref{eq:11h}, \eqref{eq:11i}, \eqref{eq:19}-\eqref{eq:22}  \label{eq:32a}.
\end{align}
\end{subequations}

We employ a penalty-based, two-tier iterative scheme for passive beamforming. In the inner loop, a convex program over the passive variables $\{\mathbf{Q_{l}}\}$ is solved, while in the outer loop the penalty parameter is updated as
$\gamma^{(j+1)}=\alpha\,\gamma^{(j)}$ with $\alpha>1$ as the iterations progress. 

The procedure stops when $\sum_{l\in\{t,r\}}\Big(\|\mathbf{Q}_{l}\|_{*}-\widehat{\mathbf{Q}}_{l}^{(j)}\Big)\le \sigma$,
with $\sigma$ a prescribed tolerance. Problem \eqref{eq:32} is a standard SDP, which can be solved by CVX.

\begin{algorithm}[!b]
\caption{Proposed AO Solution for Problem \eqref{eq:23}}
\label{alg:srocr-p2}
\begin{algorithmic}[1]
\State \textbf{Initialize:} Determine $\Delta_{0}$ and $R_{\max}$, initialize $\{\mathbf{Q_l^{(0)}}\}$ with a random matrix. Set the iteration index $r=0$.
\Repeat
\State \textbf{Initialize SROCR:} set $i=1$; choose tolerance $\epsilon$; initialize $\{\mathbf{u_k^{(0)}},\,\mathbf{\Upsilon_k^{(0)}}\}$; solve the rank-relaxed \eqref{eq:24} with $\{\mathbf{Q_l^{(0)}}\}$ to obtain $\{\mathbf W_k^{(0)}\}$; compute the principal eigenvector $\mathbf u_{k,\max}^{(0)}$ of $\mathbf W_k^{(0)}$ $\forall k\in\mathcal K$.
\State \textbf{Check:} if \eqref{eq:27} is feasible, solve \eqref{eq:27} to get $\mathbf W_k^{(i)}$ and set $\mathbf{\Upsilon_k^{(i)}}=\mathbf{\Upsilon_k^{(i-1)}}$, $\forall k$; otherwise set $\mathbf{\Upsilon_k^{(i)}}=\tfrac{1}{2}\,\mathbf{\Upsilon_k^{(i-1)}}$.
\State \textbf{Update:} $u_k^{(i)}=\min\!\left(1,\; \dfrac{\lambda_{\max}\!\big(\mathbf W_k^{(i)}\big)}{\operatorname{tr}\!\big(\mathbf W_k^{(i)}\big)}+\Upsilon_k^{(i)}\right)$, $\forall k$.
\State \textbf{Set:} $i \leftarrow i+1$.
\State \textbf{until} $u_k^{(i)}=1$, $\forall k\in\mathcal K$, and $\big|\mathbf{\xi^{(i)}}-\mathbf{\xi^{(i-1)}}\big|\le \epsilon$.
\State \quad Given $\{\mathbf{W_k^{(r+1)}}\}$, solve \eqref{eq:32} via the proposed penalty-based scheme to update $\{\mathbf{Q_l^{(r+1)}}\}$.
\State \quad \textbf{Update iteration index:} $r\leftarrow r+1$.
\Until the iterative gain of the objective function value is below a predefined threshold $\Delta_{0}>0$ or $r=R_{\max}$.
\end{algorithmic}
\end{algorithm}

\subsection{Complexity Analysis}
The overall procedure for solving \eqref{eq:23} is summarized in \textbf{Algorithm 1}. Let $\mathcal{K}$, $\mathcal{E}$, and $\mathcal{T}$ denote the numbers of IRs, ERs, and sensing targets, respectively. In the AO outer loop, problems \eqref{eq:27} and \eqref{eq:32} are solved in an alternating fashion. The overall computational complexity is governed by the number of outer loop iterations, $I_{\text{out}}$. Drawing from \cite{5447068}, the active beamforming complexity is
$\mathcal{O_{A,\text{(SR)}}}=I_{\text{SR}}\,\tilde{\mathcal O}\!\big((\mathcal{K}{+}\mathcal{E}{+}\mathcal{T})\big((MN)^{3.5}+M^{3.5}\big)\big)$
where $I_{\text{SR}}$ is the number of SROCR iterations. The passive beamforming, requiring $I_{\text{in}}$ inner iterations, has a complexity of $I_{\text{in}} \cdot \mathcal{O}_P$ with
$\mathcal{O}_P=\tilde{\mathcal O}\!\big((\mathcal{K}{+}\mathcal{E}{+}\mathcal{T})(MN)^{3.5}+2N^{3.5}\big)$.
Thus, the total complexity is
$\mathcal{O}_{\text{total}} = I_{\text{out}} \cdot \big(\mathcal{O_{A,\text{(SR)}}} + I_{\text{in}} \cdot \mathcal{O}_P \big)$.

\section{Numerical Results}

The BS and STAR-RIS are positioned at $(50,0,0)\,\mathrm{m}$ and $(0,0,0)\,\mathrm{m}$, respectively. We consider two IRs and two ERs, with the passive targets placed on a circle of radius $3\,\mathrm{m}$ on each side of the STAR-RIS. The angles of the BS--STAR paths are drawn independently from a uniform distribution. Large-scale attenuation follows $\nu = C_{0}\,(d/D_{0})^{-\alpha}$, where $C_{0}=-30\,\mathrm{dB}$ is the path loss at the reference distance $D_{0}=1\,\mathrm{m}$, $d$ is the BS--STAR separation, and $\alpha=2.2$ is the path-loss exponent. Other system parameters are set as follows: carrier frequency $f_c=10\,\text{GHz}$ with wavelength $\lambda_c=0.03\,\text{m}$, noise power $\sigma_{I}^2=\sigma_{e}^2= -110\,\text{dBm}$, number of BS antennas $M=16$, L=16, transmit power budget $P_{\max}=20\,\text{W}$, required user rate $R_{\text{th}}=2\,\text{bit/s/Hz}$, eavesdropper-rate limit $R_{e,\text{th}}=1\,\text{bit/s/Hz}$, and beampattern level $\Lambda=1\,\text{dB}$, $u_k^{(0)} = 0$,$\Upsilon_k^{(0)} = 0.1$, $\Delta_0=10^{-3}$, $\sigma=10^{-7}$, $\epsilon=10^{-4}$. For a UPA-type STAR-RIS with $5\times 8=40$ elements operating at $10\,\text{GHz}$, the Rayleigh distance is approximately $5\,\text{m}$, which places all users in the near field. The channel estimation errors for IRs, ERs, and targets are modeled as
$\rho_{\text{IR}}=\delta_{k}/\| \hat{\mathbf{H}}_k \|$, 
$\rho_{\text{E}}=\delta_{e}/\| \hat{\mathbf{F}}_e \|$, and 
$\rho_{\text{T}}=\delta_{t}/\| \hat{\mathbf{H}}_{t} \|$, respectively. 
We adopt a common uncertainty level $\rho$ such that 
$\rho=\rho_{\text{IR}}=\rho_{\text{E}}=\rho_{\text{T}}$.

\begin{figure*}
  \centering
  \subfloat[Total harvested power vs. AO iteration index.]{%
    \includegraphics[width=0.31\linewidth]{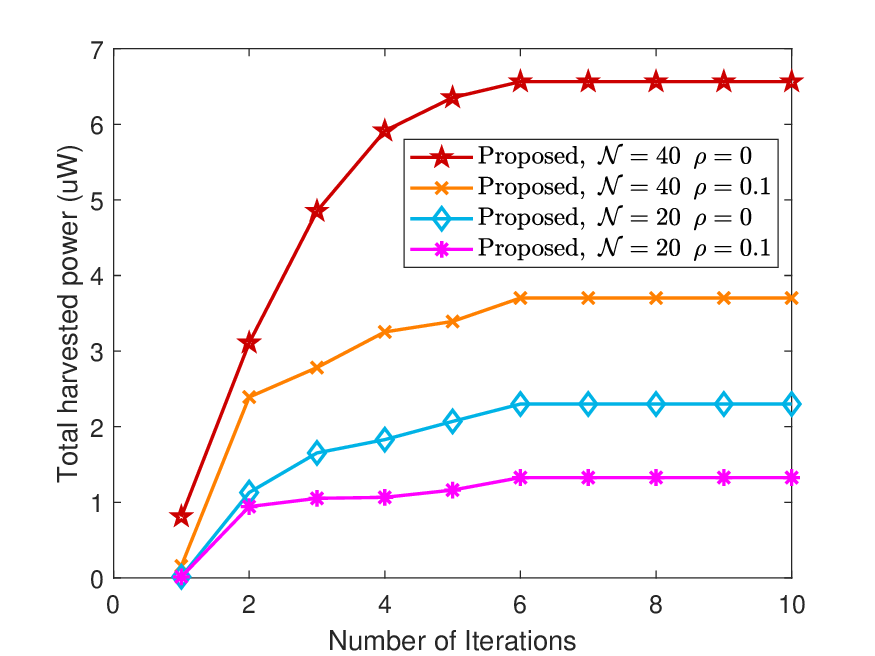}%
    \label{fig:iter}}
  \hspace{0.015\linewidth}
  \subfloat[Total harvested power vs. transmit power.]{%
    \includegraphics[width=0.31\linewidth]{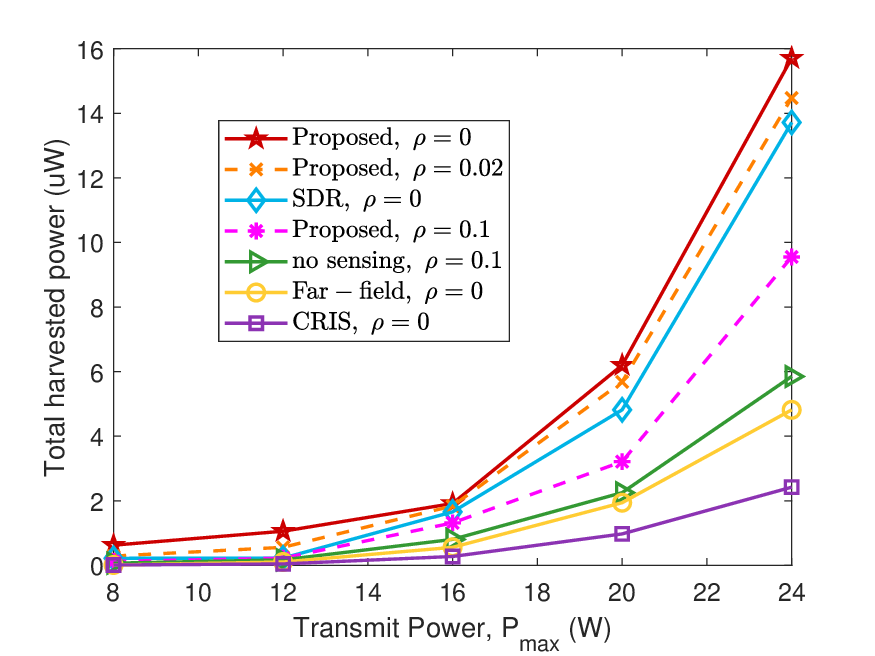}%
    \label{fig:power}}
  \hspace{0.015\linewidth}
  \subfloat[Total harvested power vs. required rate.]{%
    \includegraphics[width=0.31\linewidth]{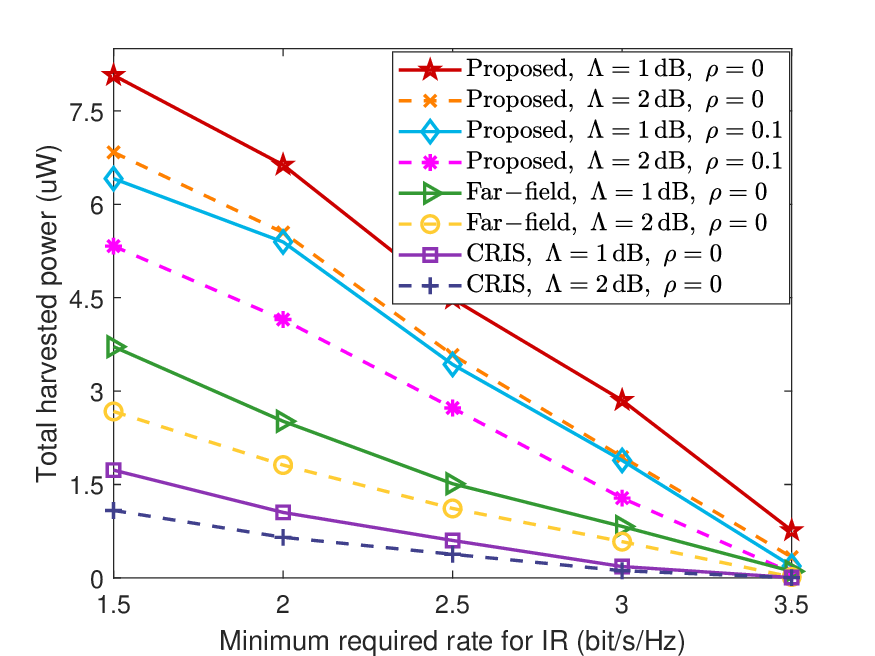}%
    \label{fig:rate}}
  \caption{Total harvested power vs. iteration index, maximum transmit power, and user rate for different beampattern-gain levels.}
  \label{fig:results}
\end{figure*}

Fig. \ref{fig:results}\subref{fig:iter}  shows the harvested power versus the number of iterations for the proposed method and baseline schemes under different channel estimation errors. All schemes reach steady convergence by the sixth iteration, confirming the effectiveness of the AO framework. Larger STAR-RIS arrays reach higher plateaus, while imperfect CSI lowers the attainable harvested power. These results emphasize the need for robust design to preserve near-field focusing gains.

Fig. \ref{fig:results}\subref{fig:power} plots the total harvested power versus different maximum transmission power. We benchmark four schemes: (i) Far-field only, where all links are modeled using a far-field channel; (ii) conventional RIS (CRIS), where the STAR-RIS is evenly split so that half the elements operate in transmission-only mode and the other half in reflection-only mode; (iii) semidefinite relaxation (SDR), which relaxes the rank-one constraint in the active-beamforming semidefinite program; and (iv) no sensing, where the BS transmits only information-bearing signals (no dedicated sensing waveform). In Fig. \ref{fig:results}\subref{fig:power}, total harvested power grows rapidly with the transmit power, and the proposed near-field design yields the highest harvested power across all $P_{max}$. Near-field beamforming surpasses far-field, and the gap widens as $P_{max}$ increases due to stronger range-dependent focusing. The proposed scheme outperforms the SDR and the no sensing scheme. SDR cannot always enforce a rank-one solution, which weakens beam focusing. The no sensing baseline highlights the benefit of joint dual-waveform design in ISCPT, where adding a radar sensing waveform provides extra degrees of freedom for energy delivery. CRIS remains below the near-field designs because fixed per-element mode selection limits beam control flexibility. Larger CSI uncertainty $(\rho=0,\ 0.02,\ 0.1)$ leads to a monotonic reduction in harvested power.

Fig. \ref{fig:results}\subref{fig:rate} displays the performance gains versus the minimum required rate for cell users. As the minimum required rate increases, total harvested power drops because more resources are steered to information beams, while the proposed near-field design remains the best. Higher sensing threshold $(\Lambda=2\,\mathrm{dB})$ and larger channel uncertainty $(\rho=0.1)$ both reduce harvested power, yet the ordering remains unchanged. Far-field baselines lag due to the loss of range-dependent focusing. CRIS is lowest because fixed per-element mode selection limits the available degrees of freedom and reduces the effective aperture gain.

\section{Conclusion}
We proposed a robust near-field STAR-RIS–aided ISCPT framework. It employs AO with the S-procedure, SROCR for rank-one active beamforming, and a penalty-based SCA update for STAR-RIS coefficients, ensuring feasibility under bounded CSI errors and meeting rate, secrecy, and beampattern constraints. Our simulation results demonstrated that near-field beamforming can significantly enhance the harvested sum power for STAR-RIS-aided ISCPT multifunction systems.




\bibliographystyle{IEEEtran}  
\bibliography{STAR_RIS_NF}

@ARTICLE{10304608,
  author={Zhu, Guangyu and Mu, Xidong and Guo, Li and Huang, Ao and Xu, Shibiao},
  journal={IEEE Transactions on Wireless Communications}, 
  title={Robust Resource Allocation for STAR-RIS Assisted SWIPT Systems}, 
  year={2024},
  volume={23},
  number={6},
  pages={5616-5631},
  doi={10.1109/TWC.2023.3327502}}

@book{boyd2004convex,
  title={Convex optimization},
  author={Boyd, Stephen and Vandenberghe, Lieven},
  year={2004},
  publisher={Cambridge university press}
}

@inproceedings{CaoEUSIPCO2017,
  author    = {P. Cao and J. Thompson and H. V. Poor},
  title     = {A sequential constraint relaxation algorithm for rank-one constrained problems},
  booktitle = {Proc. Eur. Signal Process. Conf. (EUSIPCO)},
  year      = {2017},
  pages     = {1060--1064}
}

@ARTICLE{10663809,
  author={Li, Xiaoyang and Han, Zidong and Zhu, Guangxu and Shi, Yuanming and Xu, Jie and Gong, Yi and Zhang, Qinyu and Huang, Kaibin and Letaief, Khaled B.},
  journal={IEEE Communications Magazine}, 
  title={Integrating Sensing, Communication, and Power Transfer: From Theory to Practice}, 
  year={2024},
  volume={62},
  number={9},
  pages={122-127},
  doi={10.1109/MCOM.001.2300623}}

@ARTICLE{8901159,
  author={Elayan, Hadeel and Amin, Osama and Shihada, Basem and Shubair, Raed M. and Alouini, Mohamed-Slim},
  journal={IEEE Open Journal of the Communications Society}, 
  title={Terahertz Band: The Last Piece of RF Spectrum Puzzle for Communication Systems}, 
  year={2020},
  volume={1},
  number={},
  pages={1-32},
  doi={10.1109/OJCOMS.2019.2953633}}

@ARTICLE{10716601,
  author={Liu, Yuanwei and Ouyang, Chongjun and Wang, Zhaolin and Xu, Jiaqi and Mu, Xidong and Swindlehurst, A. Lee},
  journal={IEEE Communications Surveys and Tutorials}, 
  title={Near-Field Communications: A Comprehensive Survey}, 
  year={2025},
  volume={27},
  number={3},
  pages={1687-1728},
  doi={10.1109/COMST.2024.3475884}}

@ARTICLE{11146519,
  author={Yi, Yaqian and Zhang, Guangchi and Cui, Miao and You, Changsheng and Wu, Qingqing},
  journal={IEEE Wireless Communications Letters}, 
  title={AN-Aided Secure Beamforming for ELAA-SWIPT in Mixed Near-and Far-Field}, 
  year={2025},
  volume={},
  number={},
  pages={1-1},
  doi={10.1109/LWC.2025.3605106}}

@ARTICLE{9133156,
  author={Liu, Hang and Yuan, Xiaojun and Zhang, Ying-Jun Angela},
  journal={IEEE Journal on Selected Areas in Communications}, 
  title={Matrix-Calibration-Based Cascaded Channel Estimation for Reconfigurable Intelligent Surface Assisted Multiuser MIMO}, 
  year={2020},
  volume={38},
  number={11},
  pages={2621-2636},
  doi={10.1109/JSAC.2020.3007057}}

@ARTICLE{10559261,
  author={Zhang, Zheng and Liu, Yuanwei and Wang, Zhaolin and Mu, Xidong and Chen, Jian},
  journal={IEEE Internet of Things Journal}, 
  title={Simultaneous Wireless Information and Power Transfer in Near-Field Communications}, 
  year={2024},
  volume={11},
  number={16},
  pages={27760-27774},
  doi={10.1109/JIOT.2024.3402556}}

@ARTICLE{10498098,
  author={Zhao, Boqun and Ouyang, Chongjun and Liu, Yuanwei and Zhang, Xingqi and Poor, H. Vincent},
  journal={IEEE Journal of Selected Topics in Signal Processing}, 
  title={Modeling and Analysis of Near-Field ISAC}, 
  year={2024},
  volume={18},
  number={4},
  pages={678-693},
  doi={10.1109/JSTSP.2024.3386054}}

@ARTICLE{11072251,
  author={Chen, Ziqiang and Wang, Feng and Han, Guojun and Wang, Xin and Lau, Vincent K. N.},
  journal={IEEE Wireless Communications Letters}, 
  title={Robust Beamforming Design for Secure Near-Field ISAC Systems}, 
  year={2025},
  volume={},
  number={},
  pages={1-1},
  doi={10.1109/LWC.2025.3586249}}

@INPROCEEDINGS{10437418,
  author={Li, Haochen and Liu, Yuanwei and Mu, Xidong and Chen, Yue and Zhiwen, Pan},
  booktitle={ Proc. GLOBECOM 2023 - 2023 IEEE Global Communications Conference}, 
  title={Joint Beamforming for STAR-RIS in Near-Field Communications}, 
  year={2023},
  volume={},
  number={},
  pages={625-630},
  doi={10.1109/GLOBECOM54140.2023.10437418}}

@ARTICLE{5447068,
  author={Luo, Zhi-quan and Ma, Wing-kin and So, Anthony Man-cho and Ye, Yinyu and Zhang, Shuzhong},
  journal={IEEE Signal Processing Magazine}, 
  title={Semidefinite Relaxation of Quadratic Optimization Problems}, 
  year={2010},
  volume={27},
  number={3},
  pages={20-34},
  doi={10.1109/MSP.2010.936019}}

@ARTICLE{10918637,
  author={Rostamikafaki, Zahra and Chan, Francois and D’Amours, Claude},
  journal={IEEE Access}, 
  title={Outage-Constrained Secrecy Rate Maximization for STAR-RIS With Energy-Harvesting Eavesdroppers and Imperfect CSI}, 
  year={2025},
  volume={13},
  number={},
  pages={47927-47937},
  doi={10.1109/ACCESS.2025.3549450}}

\end{document}